\documentclass[prl,twocolumn,amssymb,showpacs]{revtex4}

\usepackage{bm}
\usepackage{graphicx}
\usepackage{dcolumn}
\begin{document}

\title{Photon emission as a source of coherent behaviour of polaritons}

\author{Herbert Vinck-Posada, Boris A. Rodriguez}
\affiliation{Instituto de Fisica, Universidad de Antioquia, AA 1226, Medellin, Colombia}
\author{P.S.S. Guimaraes}
\affiliation{Universidade Federal de Minas Gerais, Departamento de Fisica, C. P. 702, Belo Horizonte, Brasil}
\author{Alejandro Cabo and Augusto Gonzalez}
\affiliation{Instituto de Cibern\'etica, Matem\'atica y F\'{\i}sica, Calle
 E 309, Vedado, Ciudad Habana, Cuba}

\begin{abstract}
We show that the combined effect of photon emission and Coulomb interactions may drive an exciton-polariton system towards a dynamical coherent state, even without phonon thermalization or any other relaxation mechanism. Exact diagonalization results for a finite  system (a multilevel quantum dot interacting with the lowest energy photon mode of a microcavity) are presented in support to this statement.
\end{abstract}

\maketitle

The effectiveness of phonon relaxation in systems of excitonic polaritons has been widely discussed recently \cite{r1} in connection to the possible Bose-Einstein condensation of polaritons \cite{ICTP2006}. Due to the very small polariton lifetime, of the order of picoseconds, and the small polariton-phonon scattering cross section \cite{PRB65(2002)165310}, there is a common belief that phonons alone can not account for thermalization of the polariton gas to the lattice temperature (the phonon bottleneck). However, the (time-resolved and angle-resolved) observed emission from decaying polaritons \cite{0604394} suggests that, for strong enough pumping pulse and a moderate positive detuning, the occupation probabilities of polariton states can be fitted to a Bose-Einstein distribution with an effective temperature even lower than the lattice temperature. The question is, thus, what is the source of coherence in such a situation?

In the present paper, we give an answer to the question above showing that phonon relaxation is not a necessary condition for the polariton system to reach a coherent state. Indeed, in our computations we show, for a finite system, that the emission of photons is also a source of coherence of polaritons leading, under certain conditions, to a high occupation probability for the ground state.

We first consider the model of paper [\onlinecite{PhysE35(2006)99}], in which a multilevel quantum dot interacts with the lowest-energy photon mode of a microcavity. The Hamiltonian describing the system is the following:

\begin{eqnarray}
H&=& \sum_{i}\left\{T^{(e)}_{i}e^\dagger_i e_i+ T^{(h)}_{\bar i}h^\dagger_{\bar i} h_{\bar i}\right\} \nonumber\\
&+& \frac{\beta}{2} \sum_{ijkl}\langle ij||kl\rangle~ e^\dagger_i e^\dagger_j e_l e_k +\frac{\beta}{2} \sum_{\bar i\bar j\bar k\bar l}\langle \bar i\bar j||\bar k\bar l\rangle~ h^\dagger_{\bar i} h^\dagger_{\bar j} h_{\bar l} h_{\bar k}
\nonumber\\
&-& \beta \sum_{i\bar j k\bar l}\langle i\bar j||k\bar l\rangle~ e^\dagger_{i} h^\dagger_{\bar j} h_{\bar l} e_{k} + (E_{gap}+\hbar\omega)~ a^\dagger a\nonumber\\
&+& g \sum_i \left\{ a^\dagger h_{\bar i}e_i+a e^\dagger_i h^\dagger_{\bar i}\right\}.
\label{eq1}
\end{eqnarray}

\noindent
We include 12 single-electron and 12 single-hole (two-dimensional, harmonic oscillator) levels in Eq.  (\ref{eq1}). We assume that both electrons and holes are confined in a small spot of a quantum well, in such a way that the single-particle spectrum is almost flat:

\begin{equation}
T^{(e)}_{i}=E_{gap}, ~~T^{(h)}_{\bar i}=0.
\end{equation}

\begin{figure}[t]
\begin{center}
\includegraphics[width=.95\linewidth,angle=0]{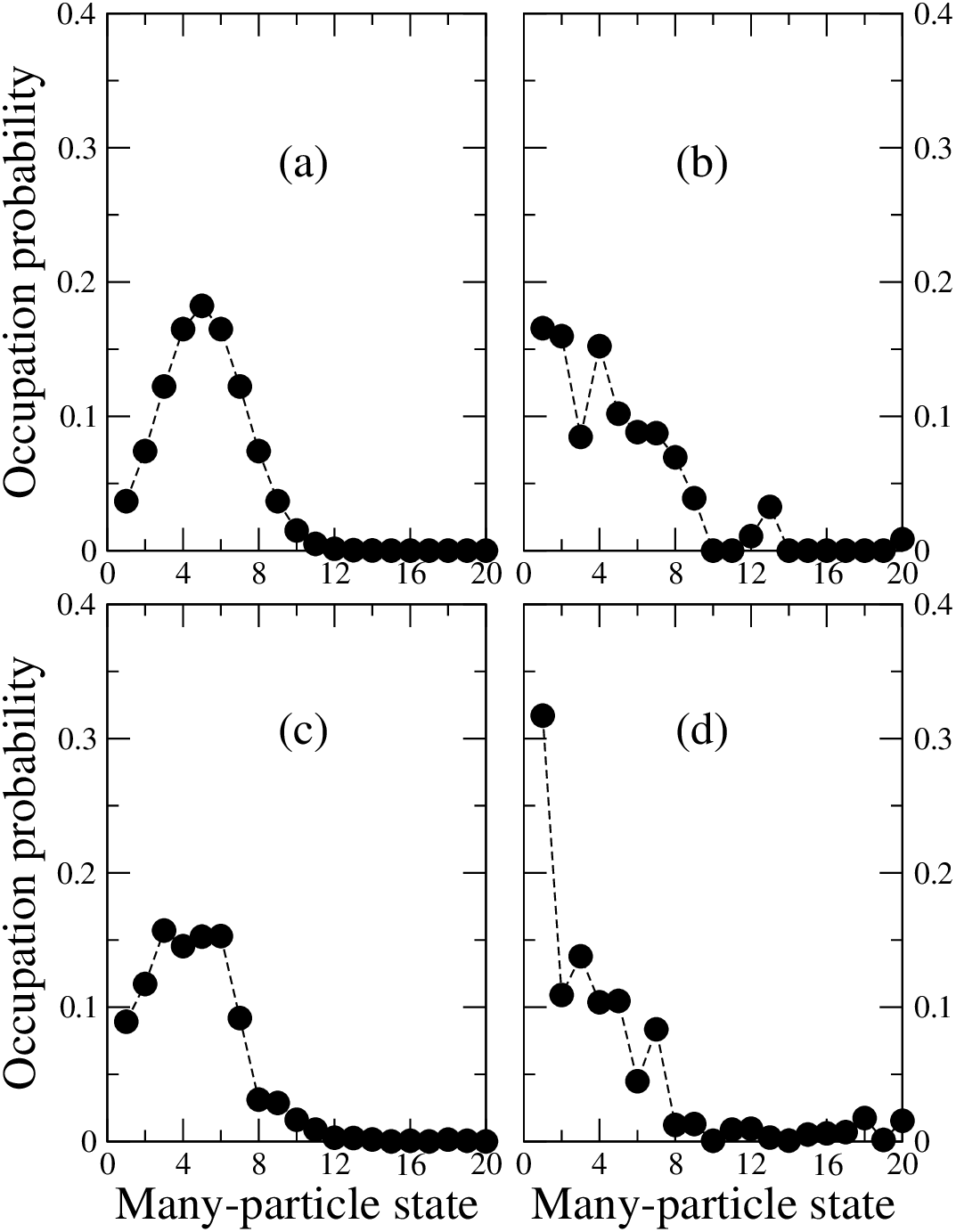}
\caption{\label{fig1} Occupation probabilities of the first 20 states in the 10-polariton system (initial distribution, panel (a)) and in the 5-polariton systems after the emission of 5 photons: (b) $\beta=0$, $\hbar\omega=5$ meV, (c) $\beta=2$ meV, $\hbar\omega=-3$ meV, (d) $\beta=2$ meV, $\hbar\omega=5$ meV.}
\end{center}
\end{figure}

\noindent
The Coulomb coupling constant, $\beta$, is modified at will in order to study its effects on the dynamics. $\langle ij||kl\rangle$ are matrix elements of the Coulomb interaction between harmonic oscillator states. The parameter $\hbar\omega$ gives the detuning of the photon energy with respect to the (bare) pair energy, and $g$ is the photon-matter coupling strength. We consider only states with total angular momentum $L=0$, which are maximally coupled to the photon mode.

The Hamiltonian, Eq. (\ref{eq1}), preserves the polariton number,

\begin{equation}
N_{pol}=a^\dagger a+\sum_i (h^\dagger_{\bar i}h_{\bar i}+ e^\dagger_i e_i)/2,
\end{equation}

\noindent
and leads to an excitation gap (the gap from the ground state to the first excited state of the system with a fixed $N_{pol}$ number) roughly proportional to $\sqrt{N_{pol}}~g$. 

Due to the small, but finite, transparency of the microcavity mirrors, photons are continuously emitted. In order to study the evolution of the system due to sucessive photon emission events, we first consider the transition probabilities:

\begin{equation}
P_{fi}\sim |\langle\psi_f|a|\psi_i\rangle|^2, 
\end{equation}

\noindent
where $\psi_i$ is a state with polariton number $N_{pol}$, and $\psi_f$ a state with $N_{pol}-1$. The probabilities should be normalized:

\begin{equation}
\sum_f P_{fi}=1.
\label{eq4}
\end{equation}

\noindent
We take an initial distribution of weights, $\rho_i^{(10)}$, for the states with $N_{pol}=10$. After the emission of one photon, the weight of the $N_{pol}=9$ state $\psi_f$ is:

\begin{equation}
\rho_f^{(9)}=\sum_i P_{fi} \rho_i^{(10)}.
\end{equation}

\noindent
Proceeding in this way, we compute $\rho_f^{(8)}$, $\rho_f^{(7)}$, etc.
This ``dynamics'' captures the basics of the Liouville equation for the density matrix and, even more, could be appropiate for the description of an experiment in which the number of photons leaving the cavity is continuously measured.

We draw in Fig. \ref{fig1} the occupation probabilities in the 5-polariton system after the emission of 5 photons. Panel (a) shows the initial distribution in the 10-polariton system. Note that the $x$-axis represents the 10-polariton states. $x=1$ refers to the ground state, $x=2$ to the first excited state, etc. Panels (b) -- (d) show the probabilities of the 5-polariton states for different choices of the parameters $\beta$ and $\hbar\omega$. The photon-matter coupling strength, $g$, was taken equal to 3 meV, corresponding to an excitation gap of around 10 meV in the 10-polariton system.

The figure shows that a positive detuning leads to an enhancement of the ground-state occupation probability. If Coulomb interactions are turned on, the ground state occupation reaches 30 \% or higher. However, for negative detuning the shape of the initial distribution is preserved. This fact, coming only from dynamical considerations, could be in the basis of the observed polariton coherence in Ref. [\onlinecite{0604394}]. 

\begin{figure}[t]
\begin{center}
\includegraphics[width=.95\linewidth,angle=0]{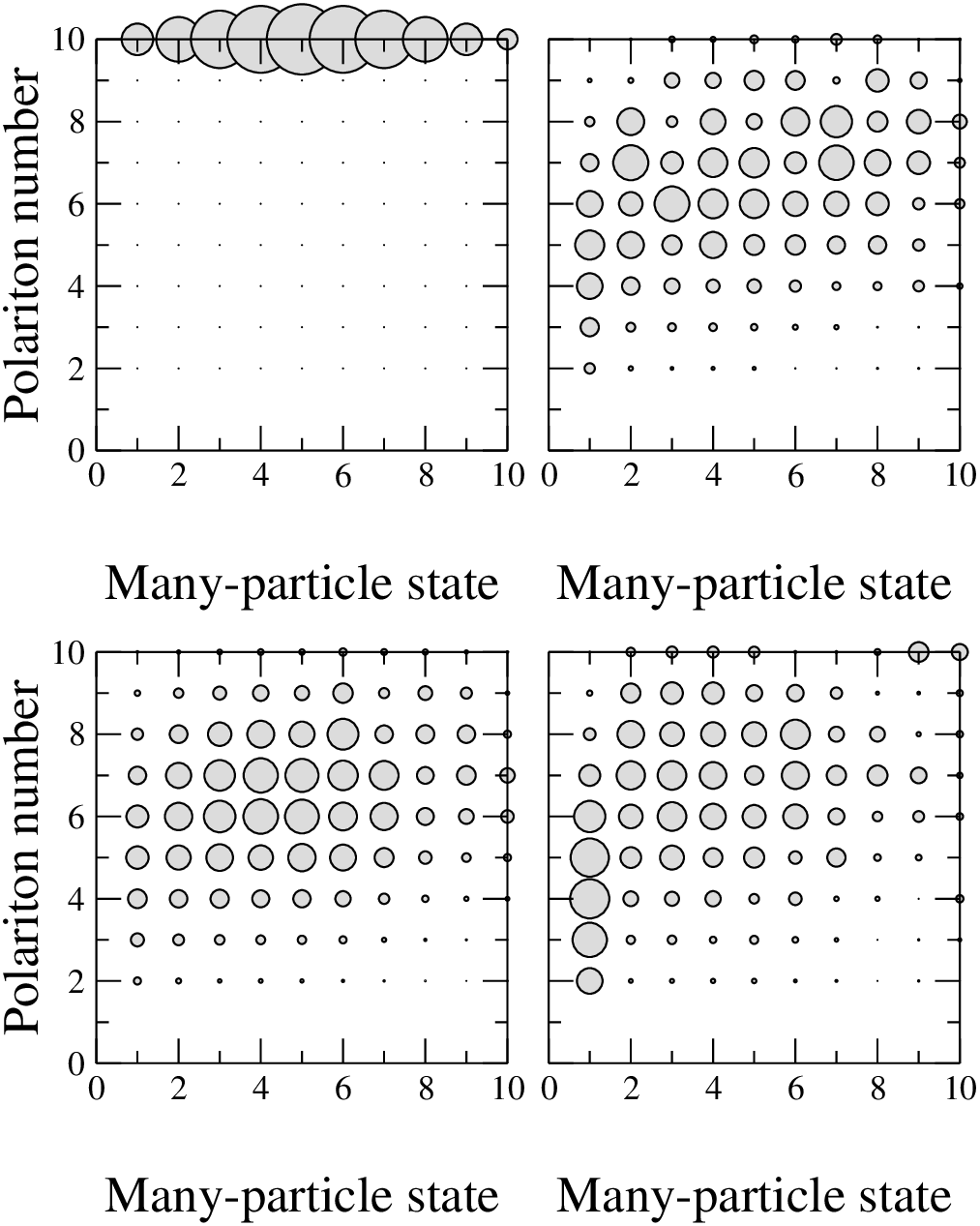}
\caption{\label{fig2} The diagonal elements $\rho_{ii}^{(Npol)}$ (occupations) as functions of $i$ ($x$ axis)
and $N_{pol}$ ($y$ axis). The areas of the circles are proportional to $\rho_{ii}^{(Npol)}$. The left upper panel corresponds to the initial distribution, whereas the other three correspond to the same sets of parameters as in Fig. \ref{fig1}. The running time is such that the mean polariton number is around 6 in each case.}
\end{center}
\end{figure}

These qualitative findings are confirmed by the numerical solution of the Liouville equation for the density matrix \cite{Tejedor,libroOC}:

\begin{eqnarray}
\frac{{\rm d}\rho_{fi}}{{\rm d}t}&=&\frac{i}{\hbar} (E_i-E_f)\rho_{fi}
+\kappa\sum_{j,k}\langle\psi_f|a|\psi_j\rangle \rho_{jk} \langle\psi_k|a^{\dagger}|\psi_i\rangle\nonumber\\
&-& \frac{\kappa}{2} \sum_{j,k} \langle\psi_f|a^{\dagger}|\psi_j\rangle
\langle\psi_j|a|\psi_k\rangle \rho_{ki}\nonumber\\
&-& \frac{\kappa}{2} \sum_{j,k} \rho_{fj} \langle\psi_j|a^{\dagger}|\psi_k\rangle \langle\psi_k|a|\psi_i\rangle, 
\end{eqnarray}

\noindent
where $\rho_{fi}=\langle \psi_f|\rho|\psi_i\rangle$. The energy eigenvalues, $E_i$, and
transition amplitudes, $\langle \psi_f|a|\psi_i\rangle$, come from the exact diagonalization of the Hamiltonian, Eq. (\ref{eq1}),  for $N_{pol}$ ranging from 1 to 10. 
The parameter $\kappa$ accounts for photon losses through the cavity mirrors ($\hbar\kappa\approx E_{gap}/Q$, where $Q$ is the cavity quality factor). In our calculations, we take $\kappa=0.1$ ps$^{-1}$. As in the previous example, the initial (mixed)
state is a distribution of weights, $\rho_{ii}^{(10)}$, for the states of the 10-polariton system. In the simulations, we take 20 states ($i=1,\dots 20$) in each sector with fixed $N_{pol}$.

A sample of the results is shown in Fig. \ref{fig2}, where the weights $\rho_{ii}^{(Npol)}$
for the state $i$ ($x$ axis) and polariton number $N_{pol}$ ($y$ axis) are represented as circles of areas proportional to $\rho_{ii}^{(Npol)}$. The distribution of panels in the figure is similar to Fig. \ref{fig1}. That is, in the left upper corner the initial ($t=0$) distribution is drawn. The mean number of polaritons is 10 at $t=0$. For the other three panels, whose $\beta$ and $\hbar\omega$ parameters are the same as in Fig. \ref{fig1}, the runing time has been chosen in such a way that the mean number of polaritons is 6.

\begin{figure}[t]
\begin{center}
\includegraphics[width=.95\linewidth,angle=0]{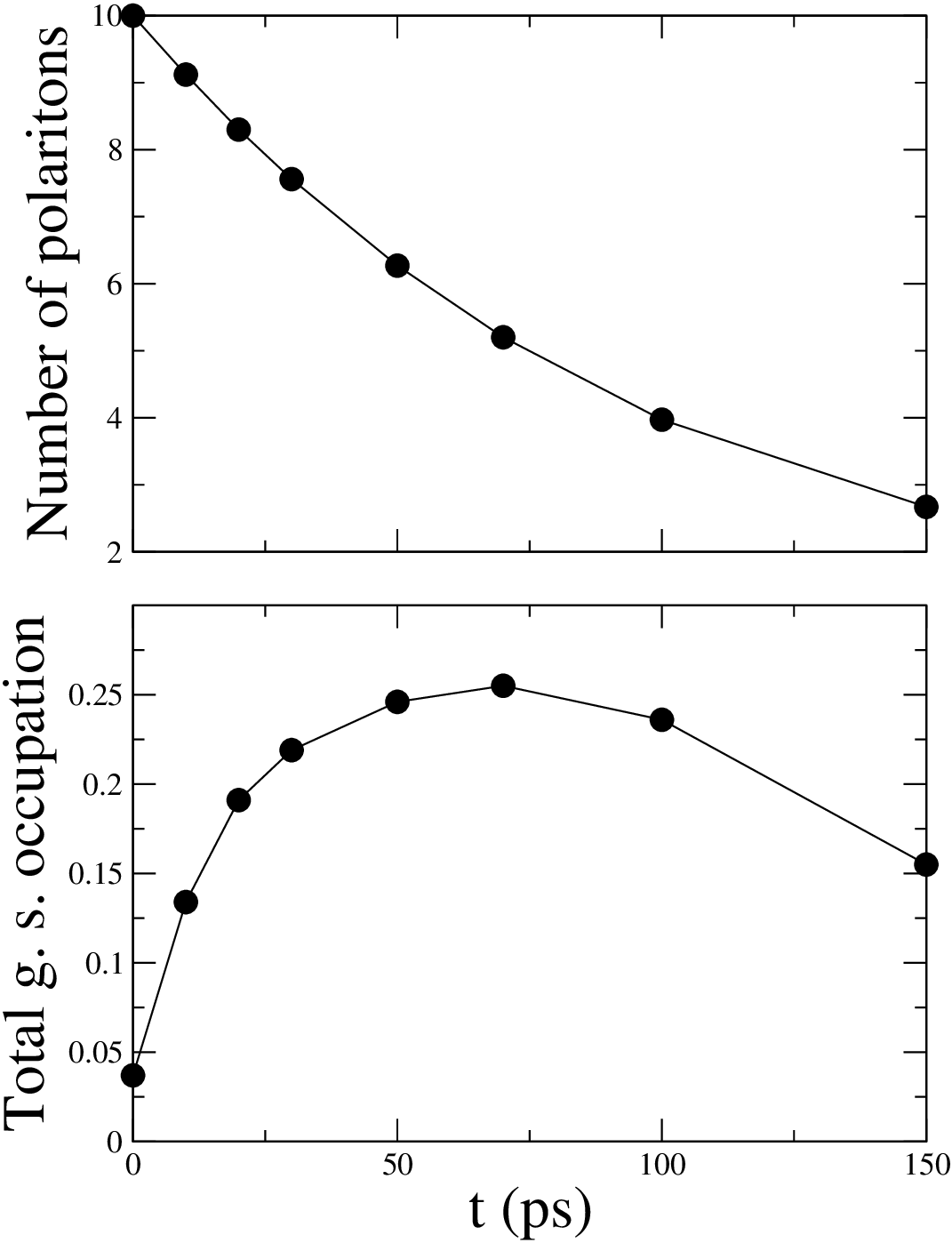}
\caption{\label{fig3} The mean number of polaritons (upper panel), and the total ground state occupation, defined in Eq. (\ref{eqx}), as functions of time. The system constants are $\hbar\omega=5$ meV, $\beta=2$ meV.}
\end{center}
\end{figure}

Besides the same qualitative feature noticed above, i.e. enhancement of ground-state occupations when both positive detuning and Coulomb interactions are combined, Fig. \ref{fig2} reveals that high ground-state occupations (as compared with the rest of the states in the same $N_{pol}$ sector) is a common characteristic of states with low polariton numbers, i.e. those who have radiated the most.

We may define a total ground-state occupation as:

\begin{equation}
\rho_{gs}^{(total)}=\sum_{Npol=1}^{10} \rho_{11}^{(Npol)}.
\label{eqx}
\end{equation}

\noindent
From our numerical solution, it follows that $\rho_{gs}^{(total)}\approx 0.25$, 0.11 and 0.07
for the sets of parameters used in Figs. \ref{fig1} and \ref{fig2}, i.e. $\beta=2$ meV, $\hbar\omega=5$ meV; $\beta=0$, $\hbar\omega=5$ meV, and $\beta=2$ meV, $\hbar\omega=-3$ meV,
respectively.

It is interesting to draw the time evolution of $\rho_{gs}^{(total)}$. This result is shown in Fig. \ref{fig3} along with the time dependence of the mean number of polaritons. The system constants are $\beta=2$ meV, $\hbar\omega=5$ meV. At $t=0$, we have $\langle N_{pol}\rangle=10$, $\rho_{gs}^{(total)}=0.036$. In around 20 ps, i.e. when $\langle N_{pol}\rangle\approx 8$ or two photons have been emitted, the total ground-state occupation rises to nearly 0.2. From $t=20$ to 70 ps ($\langle N_{pol}\rangle$ from 8 to 5) the total occupation still increases, reaching a maximum value of 0.25. Afterwards, the total occupation decreases. This later decrease is also a dynamical result, not connected with the increasing occupation of the vacuum state. Indeed, for $t=150$ ps, the weight of the vacuum state is only 0.16.

In conclusion, we have extracted energy eigenvalues and transition amplitudes from exact diagonalization calculations in order to solve the Liouville equation for the density matrix of a decaying polariton system. The results show that the emission of photons is an additional source of coherence. Further work is needed in order to conceptually clarify the role of detuning and Coulomb interactions in the dynamical coherence. 

The results should be extended to larger systems in order that their relevance to the experiment reported in Ref. \onlinecite{0604394} be more evident. Perhaps, this could be possible within the stochastic dynamics framework \cite{SSC140(2006)172}. An interesting question is whether the total ground-state occupation may reach still higher values with increasing number of particles. On the other hand, preliminary calculations based on a mean-field (BCS) dynamics \cite{PhysE27(2005)427}, similar to the semiclassical spin dynamics of the Dicke model \cite{0609169}, but including Coulomb interactions and photon losses, show that the emission of photons through leaky modes of the cavity, in spite of its incoherent nature, helps establishing a dynamical coherent state. Other mechanisms, like phonon relaxation, acting at the ``horizontal'' level in Fig. \ref{fig2}, could drive the system toward a quasiequilibrium state.

This work was partially supported by the Programa Nacional de Ciencias Basicas (Cuba), the Universidad de Antioquia Fund for Research, and the Caribbean Network for Quantum Mechanics, Particles and Fields (ICTP). A.G. acknowledges P.R. Eastham for very useful discussions.

\end{document}